\documentstyle[eqsecnum,pra,aps,preprint,epsf]{revtex}
\def\be{\begin{equation}}
\def\ee{\end{equation}}
\def\bea{\begin{eqnarray}}
\def\eea{\end{eqnarray}}
\def\pa{\partial}

\begin{document}
\draft
\begin{titlepage}
\preprint{\vbox{\baselineskip=12pt
,\rightline{IP/BBSR/99-29}
\rightline{}
\rightline{hep-th/yymmnn}}}
\title{The Holography Hypothesis in Pre-Big-Bang Cosmology with 
      String Sources}
\author{A. K. Biswas\footnote{e-mail:anindya@iopb.res.in},
J. Maharana\footnote{e-mail:maharana@iopb.res.in}}
\address{Institute Of Physics, Sachivalaya Marg,
 Bhubaneswar-751 005, India}
\date{received}
\maketitle
\begin{abstract}
The holographic ratio in Pre-big bang string cosmology is
obtained in the presence string sources. An iterative 
procedure is adopted to solve the equations of motion and 
derive the ratio in four dimensional world. 
First the zeroth order ratio is computed 
in the remote past, i.e. at $t=-\infty$, then the holographic ratio
is obtained taking into account the evolution of the 
backgrounds following the iterative procedure. The 
corrections to the zeroth order value of the ratio depends on the
form of the initial number distribution of the strings 
chosen.
Moreover, we estimate the holographic ratio in the recent past (i.e. 
when $\gamma=-\frac{1}{d}$) and in the remote past (i.e. when
$\gamma= 0$), $\gamma\equiv\frac{p}{\varrho}$, in different dimensions 
in the Einstein frame and in the string frame. 
We find that in the first case it has similar time dependences 
in both the frames,
especially in four dimensions the ratio is explicitly computed
to be the same in the two cases, whereas for $\gamma=0$ case 
the time dependence is different.
\end{abstract}
\pacs{\tt PACS number(s):~98.80.Cq,~04.30.Db}
\end{titlepage}
\section
{Introduction}
Recently, holographic principle\cite{hol} has attracted 
a lot of attention in the context of black-hole physics, 
AdS/CFT correspondence\cite{ads}
 and cosmology. It has been
recognised that theories with gravity are endowed with features different
from those in the flat space. This could be realised from the fact that
entropy of a black hole is proportional to the area of its horizon
according to Bekenstein-Hawking formula. The holography proposal states
that information for such theories reside on the boundary of the
spatial volume with one bit of information per unit Planck area. Thus, if
$S$ is the total entropy of the system enclosed in volume $V$ and $A$ is
the area of the boundary, the holographic bound is given by 
$\frac{S}{A}\leq 1$ in suitable units.
 This bound is saturated for a black hole. 
Indeed, for 
special class of black holes, in the string theory framework, the 
entropy can be computed microscopically also, hence the 
saturation of the holographic ratio for this class 
can be checked from an underlying microscopic theory.
\par
The principle of holography has been examined in the cosmological context
by Fischler and Susskind\cite{fish}. Bekenstein has examined the consequences
of the boundedness of $\frac{S}{A}$ in the cosmological scenario\cite{bek}
almost a decade ago. He 
suggested that the boundedness of the entropy could
be utilised as a constraint to circumvent cosmological singularities. 
Subsequently, there has been considerable interest
to study various properties of holography in cosmological situation
\cite{bakr,bak,dil,venezia,chol}.
It is worthwhile to mention that there have been interesting developments
to bound the entropy and invoke thermodynamical considerations in the
cosmological context. 
Recently, a Hubble entropy bound was envisaged
\cite{easther,venezia} and cosmological singularties
from the string theory view point were analysed\cite{sing}. Therefore the
holography principle and generalised second law of thermodynamics in
cosmological scenario could be used as additional constraints on 
cosmological models.
Thus, these features have stimulated
study of string cosmology from a new perspective.
In another development, adopting holography as an additional principle,
a holographic covariant description of cosmology was 
proposed and is being pursued actively
\cite{busso}.
\par
The present investigation is a sequel to our previous efforts\cite{dil}
 to study
consequences of holography in the Pre-big bang (PBB) cosmology. Since string
theory describes gravity in a natural manner it is desired that the 
theory will be able to resolve issues pertaining to physics of black-holes
as well as 
the evolution of the Universe. In the PBB scenario \cite{PBB} inflation
is recognised to be due to stringy mechanism which has no analogue
in the Einstein gravity. It is well known that decelerating, expanding (FRW)
type solution in $t>0$ region 
can be related to an accelerating and expanding solution for
negative $t$ through scale-factor duality(SFD) and time inversion.
Thus, the accelerating power law expansion is driven by the kinetic energy  
term of the dilaton towards singularity whereas the 
decelerating, expanding (FRW) type solution for $t>0$ has singularity
in the past. This is the scenario in the so called string frame
metric. It is
proposed that a cold, flat, weakly coupled Universe proceeds towards
a hot, curved and strongly coupled phase and then it goes through 
graceful exit to the FRW-like phase.
\par
In the weak coupling approximation, the tree level string effective
action, in cosmological scenario, can be used to describe the evolution
of scale-factor, dilaton and other matter fields. However, as one
approaches the high curvature, strong coupling regime, this approximation
is unlikely to hold. Therefore, when one approaches $t\rightarrow 0_{-}$,
it is necessary to account for the higher order correction in $\alpha^{'}$
as well as higher genus correction. 
\par
There have been attempts 
to study cosmological evolution of graviton and
dilaton in the presence of classical stringy matter
source by several authors. For an early account in the context of 
PBB scenario 
we refer the reader 
to ref.\cite{PBBISC}. In the string theory,
this stringy matter source is taken care of 
by a phenomenological source term in the 
string effective action. 
The dynamical equation of such extended objects have interesting features
in the presence of time-dependent metric, especially if there is a horizon.
When these objects are well within the horizon, the ratio of pressure and
the energy density denoted as $\gamma$ is zero and the evolution equation
is described by the motion of the center of mass of the string and the
oscillatory terms. On the other hand, 
if a string crosses the horizon its dynamical
degrees of freedom gets frozen,
 it increases in size linearly with
time and triggers Jean's like instability \cite{cons}. Then it is termed
unstable string. When all the strings exceed the size of the horizon
in the $(1+d)$-dimension, $\gamma$ becomes $-\frac{1}{d}$.
Here, in this paper  
we will be studying the effect of such stringy sources on the
holographic properties in the cosmological context. We shall confine 
our attention on the PBB scenario, particularly in the PBB phase.
We will see that the effects 
of this stringy matter sources 
are felt reasonably only for negative time 
far away from the singularity.
\newline
The organisation of the paper is as follows:
\newline 
In the section II, we review in detail the isotropic, homogeneous 
solutions for $\gamma=0$ and $\gamma=-\frac{1}{3}$.
$\gamma$ is zero in the far past (i.e. at $t=-\infty$) and the second case 
occurs in the recent past i.e. in the vicinity of $-t_{c}$ (see Fig.1).
\newline 
Then in the section III, 
we discuss about the models of initial distribution 
of string sources, consider the relevent iteration procedure to obtain
background field configurations
from the zeroth order solutions 
and derive the form of the
the holographic ratio that we will be using in the later sections. 
Subsequently, we use this form to estimate the holographic 
ratio in the zeroth order and discuss the features
associated with finite value of the ratio.
\newline
In the next section IV, we compute the 
corrections to the holographic ratio for the power law and exponential 
distributions and study their properties for the two cases.
\newline 
In the section V, we
estimate the holographic ratio when all the strings cross the horizon
in both the string and Einstein frames in four dimensional world
and explain in some detail the known\cite{MPL}
 procedure of going to the Einstein 
frame.
\newline
In the following section VI, we deal with the
holographic ratios in the general $D$-dimension in both the remote and
recent past in the string as well as in the
Einstein frames.
\newline
We end with a discussion in the section VI.
\section
{Isotropic and homogeneous solutions in the String Frame}
The low energy effective action in the four dimension in the string frame
 is given by \cite{mov}
\be
\hbar^{-1}S^{s}=-\frac{1}{2 l_{s}^{2}}
      \int{d^{4}x \sqrt{-g} e^{-\phi}(R+\partial_{\mu}\phi \partial^{\mu}\phi)}
      +S_{\sigma}
\ee
where, $l_{s}$ is the string scale, $R$ is the curvature scalar 
computed in the string frame metric. $\phi$ is the dilaton.
The string coupling constant, $g_{s}$, is defined by the relation 
$g_{s}=e^{\frac{\phi}{2}}$. The last term in eqn.(2.1) is due to the 
string source and its effects are treated classically.
The string source part of the action is 
\be
S_{\sigma}=\frac{1}{4\pi\alpha^{'}}
           \int{d^{2}\sigma \partial_{\alpha}X^{\mu}
           \partial^{\alpha}X^{\nu}g_{\mu\nu}}
\ee
The corresponding equations of motions are
\bea
2(R_{\mu}^{\nu}+\nabla_{\mu}\nabla^{\nu}\phi)=2l_{s}^{2}e^{\phi}
T_{\mu}^{\nu},\\
R-(\nabla_{\mu}\phi)^{2}+2\nabla_{\mu}\nabla^{\mu}\phi=0
\eea
where,
\be
T_{\mu\nu}=\frac{2}{\sqrt{g}}\frac{\delta S_{\sigma}}{\delta g^{\mu\nu}}
\ee
is the stress-energy-momentum tensor following 
the definition of ref.\cite{Birrel}.

We work in isotropic and homogeneous space, therefore the line element 
in the string frame metric is
\be
ds^{2}=dt^{2}-a^{2}(dx^{i})^{2}
\ee
The energy and pressure are defined as follows
 \be
T_{\mu}^{\mu}=(T_{0}^{0},T_{i}^{i})=(\varrho,-p,-p,-p)
\ee
Then equations of motion in time-dependent form \cite{PBBISC,NET} are
\bea
\dot{\bar{\phi}}^{2}-2\ddot{\bar{\phi}}+3H^{2}=0\\
\dot{\bar{\phi}}^{2}-3H^{2}=2 l_{s}^{2}\bar{\varrho}e^{\bar{\phi}}\\
2(\dot{H}-H\dot{\bar{\phi}})=2 \l_{s}^{2}\bar{p}e^{\bar{\phi}}
\eea
where, the SFD invariant variables are defined as 
\be
\bar{\phi}=\phi-3\ln{a},\quad \bar{\varrho}=\varrho\sqrt{|g|},\quad
\bar{p}=p\sqrt{|g|}=\gamma\bar{\varrho}
\ee
and $\bar{\phi}$ is the shifted dilaton. 
The covariant conservation of stress-energy-momentum tensor takes the form
\be
\dot{\bar{\varrho}}+3H\bar{p}=0
\ee
Let us now introduce  
a new dimensionless time parameter $x$,
such that
$2l_{s}^{2}\bar{\varrho}=\frac{1}{l}\frac{dx}{dt}$;
and define $\Gamma$ by
$\gamma(x)=\frac{d\Gamma}{dx}$. 
Eqn.s (2.8)-(2.10) on integration 
reduce to\cite
{PBBISC,DPISC,MPL,NET}
\bea
\frac{d\ln{a}}{dx}&=&\frac{2\Gamma}{(x+x_{0})^{2}-3\Gamma^2}\\
\frac{d\bar{\phi}}{dx}&=&-\frac{2(x+x_{0})}{(x+x_{0})^{2}-3\Gamma^2}\\
2l_{s}^{2}\bar{\varrho}&=&
\frac{e^{\bar{\phi}}}{4l^{2}}[(x+x_{0})^{2}-3\Gamma^2]
\eea
Note that $\Gamma=\gamma x +X$ for constant $\gamma$, where
$X$ and $x_{0}$ are constants of integration. 
Obviously, $x_{0}$ is chosen 
such that $\bar{\phi}$ reaches its maximum at $x=x_{0}$.
Again, this set of first order eqn.s(2.13)-(2.15)
 can be integrated for constant
$\gamma$ to give\cite{PBBISC,DPISC,MPL} 
\bea
a&=&a_{0}|(x-x_{+})(x-x_{-})|^{\frac{\gamma}{\alpha}}
|\frac{x-x_{+}}{x-x_{-}}|^{\tilde{\alpha}}\nonumber\\
e^{\bar{\phi}}&=&e^{\bar{\phi_{0}}}
|(x-x_{+})(x-x_{-})|^{-\frac{1}{\alpha}}
|\frac{x-x_{+}}{x-x_{-}}|^{-\sigma}\nonumber\\
2l_{s}^{2}\bar{\varrho}&=&\frac{\alpha}{4l^{2}}e^{\bar{\phi_{0}}}
|(x-x_{+})(x-x_{-})|^{\frac{\alpha-1}{\alpha}}
|\frac{x-x_{+}}{x-x_{-}}|^{-\sigma}
\eea
where,
\be
\alpha=1-3\gamma^{2},\quad\sigma=\sqrt{3}\frac{\gamma}{\alpha},
\quad\tilde{\alpha}=\frac{1}{\sqrt{3}\alpha}
\ee
and
\be
x_{\pm}=\frac{1}{\alpha}[\sqrt{3}X(\sqrt{3}\gamma\pm{1})-x_{0}
(1\pm\sqrt{3}\gamma)]
\ee

Now, in the remote past, $\gamma=0$, so $\Gamma = X$.
Let us take $\gamma=0$ to start with and examine the consequences.
Then we will get the zeroth order solutions \cite{DPISC}, 
which are given below: 
\bea
a&=&a_{0}|\frac{x-x_{+}}{x-x_{-}}|^{\frac{1}{\sqrt{3}}}\\
e^{\bar{\phi}}&=&e^{\phi_{0}}|(x-x_{+})(x-x_{-})|^{-1}\\
2l_{s}^{2}\bar{\varrho}&=&\frac{e^{\phi_{0}}}{4l^{2}}
\eea
with, $x_{\pm}=\pm{\sqrt{3}X}-x_{0}$.\\
Setting
\be
x_{-}=0,\quad x_{0}=-\frac{e^{\phi_{0}}}{4l}T,
\ee
one arrives at\cite{DPISC,NET}
\bea
a&=&a_{0}(1-\frac{2T}{t})^{\frac{1}{\sqrt{3}}}\\
e^{\bar{\phi}}&=&\frac{16 l^{2} e^{-\phi_{0}}}{|t(t-2T)|}\\
2 l_{s}^{2}\bar{\varrho}&=&\frac{e^{\phi_{0}}}{4l^{2}}\\
X&=&\Gamma^{0}=\frac{e^{\phi_{0}}}{4\sqrt{3}l}T
\eea
As the Universe crosses the time
$-T$, kinetic energy of the dilaton and the curvature becomes 
comparable to the source energy density i.e. 
$\dot{\bar{\phi}}^{2} \sim H^{2} \sim \bar{\varrho}e^{\bar{\phi}}$.
We assume the source energy density per unit comoving volume to be 
small\cite{DPISC,NET}. 
So it does not
affect the initial curvature of the Universe.
We note that the solutions(2.23)-(2.25) go over to the 
dilaton driven vaccuum solution in the
$t\rightarrow 0_{-}$ limit.
Moreover, we will see that this solution acts as good string perturbative
vaccuum. In other words, 
corrections to zeroth order solutions 
is rather small. This observation is valid,
 at least, for very large negative $t$.
Let us proceed to discuss about the corrections
in the presence of
classical string sources. The length of 
a string will vary, in principle, from $l_{s}$ to $\infty$. 
At the time $t=-\infty$, horizon is
of infinite extent. All the strings are within the 
horizon. As time progresses, the horizon shrinks. So strings
also start crossing the horizon, making pressure negative and
$\gamma$ non-zero. This non-zero $\gamma$ 
will introduce corrections on the top 
of the zeroth order solution.
Moreover, $\gamma$ will be small at the
beginning. 

\begin{figure}[h]
\centerline{\leavevmode\epsfysize=2.5truecm \epsfbox{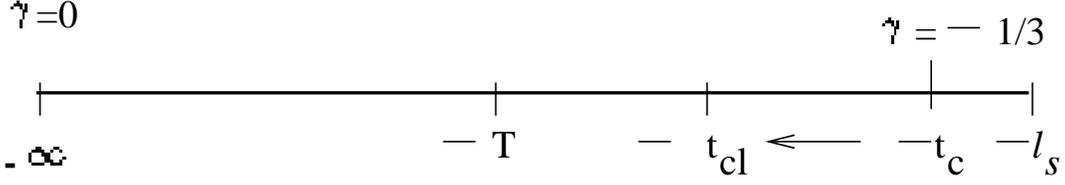}}
\caption{Temporal History Of The String-driven Pre-big-bang Phase}
\label{a}
\end{figure}
Untill now we have been discussing solutions in one limit.
Now let us consider the background field configurations 
in another limit, namely when all the 
strings are outside the horizon and
$\gamma$ is $-\frac{1}{3}$. Let $-t_{c}$ be the 
time when $\gamma$ approaches $-\frac{1}{3}$.
 Then $t_{c}$ is $l_{s}$ 
when the mean length of the strings is of the order of  $l_{s}$. 
If the mean length of strings is more, $t_{c}$
has higher value. 
Now the critical density parameter \cite{DPISC,MPL}
$\Omega(x)=\frac{\varrho e^{\phi}}{6 H^{2}}$ has the following expression,
\be
\Omega(x)
=\frac{|(x-x_{+})(x-x_{-})|}{(x-3X)^{2}}
\ee
So at $x=x_{\pm}$ or, $t\sim l_{s}$, $\Omega(x)$ goes to zero. 
Then the string sources
become unimportant compared to curvature. The tree level
string effective action is not reliable in this regime. 
Hence we consider the large
$|x|$ limit i.e. $|x|>>x_{\pm}$, then $|t|>>l_{s}$. Let us denote the
corresponding $x-time$ and the ordinary time by
$x_{cl}$ and $-t_{cl}$, where the tree level action is reliable. 
But at $t=-t_{cl}$, $\gamma$ need not coincide with $-\frac{1}{3}$. 
$\gamma=-\frac{1}{3}$ if $t_{cl}=t_{c}$. This situation is represented
by an arrow in the Fig.1.
\newline
Now let us assume that the mean length of our string sources
is of the order of $|t_{cl}|$.
Let us also
assume that the source energy density to be a reasonable fraction of 
curvature at that time. 
The corresponding field configurations
around $x_{cl}$ are \cite{DPISC,MPL}
\bea
a&=&a_{0}|(x-x_{+})(x-x_{-})|^{-\frac{1}{2}}
|\frac{x-x_{+}}{x-x_{-}}|^{\frac{\sqrt{3}}{2}}\nonumber\\
e^{\bar{\phi}}&=&e^{\bar{\phi_{0}}}
|(x-x_{+})(x-x_{-})|^{-\frac{3}{2}}
|\frac{x-x_{+}}{x-x_{-}}|^{\frac{\sqrt{3}}{2}}\nonumber\\
2l_{s}^{2}\bar{\varrho}&=&\frac{\alpha}{4l^{2}}e^{\bar{\phi_{0}}}
|(x-x_{+})(x-x_{-})|^{-\frac{1}{2}}
|\frac{x-x_{+}}{x-x_{-}}|^{\frac{\sqrt{3}}{2}}
\eea
where,
\be
x_{\pm}=\frac{3}{2}[X(-1\pm{\sqrt{3}})-x_{0}(1\mp{\frac{1}{\sqrt{3}}})]
\ee
Now we set $x_{-}=0$, keep the leading contributions of $x$ only and
express the background field configurations in terms of ordinary(cosmic) time.
Then the solutions become \cite{PBBISC,DPISC,MPL}
\bea
a_{s}(t)=(\frac{-t}{t_{0}})^{-\frac{1}{2}},\quad
\phi=\phi_{0n}-3\ln(\frac{-t}{t_{0}}),\quad
\varrho=-3p =\varrho_{0}(\frac{-t}{t_{0}})
\eea
with,
\be
t_{0}=a_{0}^{2}(\frac{e^{\phi_{0}}}{3l})^{-1},\quad
\phi_{0n}=\phi_{0}-3\ln{a_{0}},\quad 
\varrho_{0}=\frac{e^{\phi_{0}}}{6l^{2} a_{0}}\frac{1}{2 l_{s}^{2}}.
\ee
The constants in the eqn.(2.31) are related as
\be
\varrho_{0}e^{\phi_{0n}}t_{0}^{2}=\frac{3}{4 l_{s}^{2}}
\ee
So far we have considered solutions in two cases: (i)$\gamma=0$ and
(ii)$\gamma=-\frac{1}{3}$.
It is important to examine the background field configurations when $\gamma$
varies between these two extrema. This will be dealt with
 in the following sections.
 
\section{Classical string sources, iteration and holographic ratio} 
In this section we look for a 
simultaneous solution of the background equations of motion and
equations of motion for the string sources as $\gamma$ decreases from
zero to $-\frac{1}{3}$ with time proceeding from $-\infty$ to
$-t_{cl}$.
The nature of the evolution equations are such that we cannot get an 
analytic exact solution for 
$0>\gamma>-\frac{1}{3}$. 
Therefore we start from the $\gamma=0$ end.
There we have the exact solution for the background fields. 
If we put back this
solution in the sigma model action to evaluate $\gamma$, we recover the
value $\gamma=0$ \cite{cons},
only in the limiting case $t\rightarrow -\infty$ which is
same as $a\rightarrow a_{0}$. 
In other words, our exact solution is not simultaneous solution 
of background as well as string equations of motions for time not
exactly $-\infty$. Moreover, we cannot dispense with
 the position coordinates in the expression for pressure and 
energy density. To get a simple expression for $\gamma$
as a function of time 
we assume, following Gasperini etal \cite{NET},
that the source term represents an ensemble of classical strings. 
Initially all the strings are within the horizon, hence pressure and $\gamma$
are zero. The two types of initial 
distribution of the lengths
of the strings, we consider, will give different expressions
for $\gamma$. Here, first we review the basic elements of the
model, then derive the expression for $\gamma$ from the model to be used 
in the iteration procedure, to obtain the 
background fields. Then we utilize these solutions to evaluate the 
corresponding holographic ratios and subsequently compare and 
contrast the features
of the two types of distributions.
\par
Now let us briefly recapitulate 
some of the essential features of 
the model. For further details we refer the 
interested reader to the ref.\cite{NET}.
 We consider an ensemble of large number, say N, 
of classical strings. Moreover, let the length of 
i-th string at an instant of time $t$ be $L^{i}(t)$, 
and number density of strings of length $L$ at that instant be $n(L, t)$. 
Then $N=\int_{l_{s}}^{\infty}n(L, t)dL$.
 Here, the differential equations satisfied by $L^{i}(t)$ and $n(L, t)$ are,
\bea
\dot{L}^{i}(t)=H(t)L^{i}(t)\theta(L^{i}(t)-D(t))\nonumber\\ 
a\pa_a{n}+\pa_{L}\left[n(L,a)L\theta(L-D)\right]=0
\eea
where, $D=H^{-1}$ is the Hubble length and $H$ is the Hubble parameter.  
The energy density of stable and unstable strings (i.e. length
less or greater than the horizon) are given by 
\bea
E_{s}=\int d^{3}x \sqrt{g}\rho_{s}
 =\frac{1}{\pi \alpha^{\prime}}\int Ln(L,a)\theta(D-L)dL\\
E_{u}=\int d^{3}x\sqrt{g}\rho_{u}
=\frac{1}{\pi \alpha^{\prime}}\int Ln(L,a)\theta(L-D)dL
\eea
Note that in the far past when all the strings lie within the
horizon, $\gamma=0$ and $E_{u}=0$. On the otherhand, $\gamma=-\frac{1}{3}$
corresponds to the situation $E_{s}=0$.
So, $\gamma$ could be approximated by
\be
\gamma(t)=-\frac{1}{3}\frac{E_{u}}{E_{u}+E_{s}}
\ee
From the definition(3.3) it follows that
$\tilde{E}_{u}= \pi \alpha^{\prime}E_{u}$ satisfies a
differential equation
\be
\pa_{D}{\tilde{E}_{u}}-\frac{1}{D}\frac{\pa{\ln{a}}}{\pa{\ln{D}}}
\tilde{E}_{u}=
-D n_{-\infty}(D)
\ee
where, $n_{-\infty}(D)=n(D,t=-\infty)$ expresses the 
number density of strings of size of the horizon at time $t$ according
to the distribution of number density at the far past. Note that this
is the differential equation for evaluating $\tilde{E}_{u}$ at
any time and
we will be using it again and again. On the contrary, 
once we assume the 
form of the distribution of number density at the remote past,
$E_{s}$ can be
directly evaluated from the definition(3.2). So
let us discuss the two types of distributions,
\bea
n_{-\infty}(L)&=&\Lambda^{2}L^{-3},\\
n_{-\infty}(L)&=&\frac{N'}{L_{0}}exp(-\frac{L}{L_0})
\eea
where, $\Lambda$ and $N'$ are related to the total number of strings.
The mean length of a string in the two cases are respectively $2l_{s}$
and $l_{s}+L_{0}$.

Now for the first type of number distribution 
$\gamma$ goes to $-\frac{1}{3}$ as $t_{c}$ is of the order 
$l_{s}$ whereas for the 
second type $t_{c}$ can vary.
Let us see how this happens a la Gasperini etal \cite{NET}. First we note
that when $\gamma$ becomes $-\frac{1}{3}$ or, almost all strings
cross the horizon, $a(-t)\sim (-t)^{\alpha}$\cite{DPISC}, 
$\alpha$ being some
negative fraction. 
Then we find 
\bea
\tilde{E}_{u}&=&\Lambda^{2}\frac{H}{(1+\alpha)},\nonumber\\
\tilde{E}_{s}&=&\Lambda^{2}[(l_s)^{-1}-H],\nonumber\\
\gamma&=&-\frac{1}{3}\frac{1}{\frac{1+\alpha}{Hl_{s}}-\alpha}
\eea
So, $\gamma$ tends to $-\frac{1}{3}$ only when $D\sim l_{s}$.
\newline
Let us now consider the exponential distribution. 
The differential equation for $\tilde{E}_{u}$ near
$\gamma=-\frac{1}{3}$ yields,
\be
\tilde{E}_{u}=N'L_0
(\frac{D}{L_0})^{\alpha}\Gamma\left[2-\alpha,\frac{D}{L_0}\right]
\ee
whereas,
\be
\tilde{E}_{s}=N'L_{0}[(1+\frac{l_s}{L_0})exp(-\frac{l_s}{L_0})-
(1+\frac{D}{L_0})exp(-\frac{D}{L_{0}})]
\ee
and as a result,
\be
\gamma=-\frac{1}{3}
\frac{(\frac{D}{L_0})^{\alpha}\Gamma\left[2-\alpha,\frac{D}{L_0}\right]}
{[(1+\frac{l_s}{L_0})exp(-\frac{l_s}{L_0})-
(1+\frac{D}{L_0})exp(-\frac{D}{L_{0}})]
+(\frac{D}{L_0})^{\alpha}\Gamma\left[2-\alpha,\frac{D}{L_0}\right]}
\ee
where, $\Gamma\left[2-\alpha,\frac{D}{L_0}\right]$ is incomplete
gamma function.
 So if $L_0$ is such that it is greater than $l_s$ as well as
$t_{cl}$ (i.e. $L_{0}>l_{s}$ and $L_{0}> t_{cl}$)
then $\gamma$ becomes $-\frac{1}{3}$ at the time $t_{cl}$
 for a particular value of $L_0$, say $\frac{2f_{m}}{\sqrt{3}}t_{cl}$.
On the contrary, if $L_{0} \sim l_{s}$ then 
$\gamma$ becomes $-\frac{1}{3}$ when $t_{c}\sim l_{s}$. 
We note here that for
the exponential distribution of mean length $l_{s}$, $\gamma$ falls off
faster than that of power law.\\
Again as we are considering the situation almost in the far past
i.e. near $t\rightarrow-\infty$ where the horizon
is very large, almost all strings of any size are within the horizon. 
It should not matter much whether we take strings of mean length
$l_{s}$ or, larger and we take power law distribution of number of
strings or, exponential distribution. We will see how far this is true
in the following. Before we move on to estimate the holographic
ratio, let us describe 
the way we calculate the
background fields, the prescriptions for the 
iterations and how we put the holographic ratio
in proper form to compute it at each stage of iteration.
\par
In order to obtain the background field configurations order by order
through iteration, it is
suitable to use a varaible
$Y=\ln{\frac{a}{a_{0}}}$, instead of $x$, along the line of
ref.\cite{NET}.
 In terms
of the variable $Y$,
eqn.s (2.13)-(2.15) are rewritten as
\bea
2 l_{s}^{2}\bar{\varrho}(Y)=(\frac{\Gamma}{l})^{2} e^{\phi_{0}}
[(w^{\prime})^{2}-\frac{3}{4}w^{2}]\\
\bar{\phi}(Y)=\phi_{0}+2\ln{w(Y)}\\
w^{\prime\prime}+\frac{\Gamma^{\prime}}{\Gamma}w^{\prime}-\frac{3}{4}w=0
\eea
where, $w(Y)$ is an auxiliary function introduced through the
relation
\be
x+x_{0}=-2\Gamma\frac{w^{\prime}}{w}
\ee
Our main aim will be to solve the $w$ equation using the expression 
for $\Gamma$.
Note that  
putting $\gamma=0$ or, $\Gamma=\Gamma^{0}$ in the differential
equation for $w$ we get
\be
w=\frac{4l}{T}e^{-\phi_{0}}sinh{\frac{\sqrt{3}}{2}Y}
\ee
We use the expression(3.16) for $w$ to
find the first order corrected $\gamma$. Moreover,
for nonconstant $\gamma$, $\Gamma$ is given by,
\be
\Gamma=\Gamma^{0}+\int_{0}^{Y}\gamma(x)\frac{dx}{dY}dY.
\ee 

Let us now explain the procedure for iteration as we carry it out 
for power law distribution. 
In that case, putting the zeroth order expression for $w$ we get $\gamma$
which involves
the parameter $\epsilon$. 
Once we put the first order corrected $\gamma$ 
in the differential equation we obtain
$w$ corrected to order $\epsilon$.
This procedure is repeated to get higher order corrections in $w$.
For exponential distribution procedure is similar. Due to the 
presence of $Exp(-\frac{1}{f Y^{2}})$, there is no analog of the parameter
$\epsilon$;
though at the first sight 
it might appear that $\frac{1}{f}$ or, $f$ can be used as expansion
parameter depending on whether $f$
is greater or less than one. There, improvement on the
zeroth order result through iteration is relevent
in the powers of $Y$ only. 
Whatever be the distribution,
at each stage we
utilize the $w$ to evalute the holographic ratio. To achieve that we
write the holographic ratio in a proper form. Moreover, 
let us keep in mind that
we are considering the Hubble horizon throughout
( We mention in passing that in the remote past the event horizon goes
as square root of the Hubble horizon and for small time both converge).
Now in the string frame, $Planck length$, $l_p(t)$, is time-dependent
and is given by\cite{dine}
\be
l_p(t)=\frac{l_s}{\sqrt{\frac{V_6}{l_{s}^{6}}}}exp(\phi/2)=l_s exp(\phi/2)
\ee
We have taken
$\frac{V_6}{l_{s}^{6}}$ to be one in
eqn(1.1).
The ratio of entropy contained within the Hubble Horizon to
the horizon area is given by
\bea
l_p^{2}(t)\frac{S}{A}&=&l_s^{2}
exp(\bar{\phi})\sqrt{g}\frac{S}{A}\nonumber\\
&=&l_s^{2}
exp(\bar{\phi})
\frac{\bar{\varrho}(1+\gamma)}{T^{\prime}}
\frac{V_{H}}{A_{H}}\nonumber\\
&=&\l_s^{2}
exp(\bar{\phi})
\frac{\bar{\varrho}(1+\gamma)}{T^{\prime}}\frac{D}{3}\nonumber\\
&=&l_s^{2}
exp(\phi_0)
\frac{\bar{\varrho}(1+\gamma)}{T^{\prime}}
\frac{w(Y)^{2}}{3\frac{dY}{dt}}\nonumber
\eea
where, $T^{\prime}$ is defined to be the temperature.
It follows from the covariant conservation of $T_{\mu\nu}$
in the string frame
that the time development of string sources is adiabatic. 
So $\gamma$ changes with time 
keeping entropy per comoving volume constant.
Consequently,
\bea
l_p^{2}(t)\frac{S}{A}&=&l_s^{2}
exp(\phi_0)
\frac{\bar{\varrho}_{ind}f(Y)}{T^{'}_{ind}f(Y)}
\frac{w^2}{3\frac{dY}{dt}}\nonumber\\
&=&l_s^{2}
exp(\phi_0)
\frac{exp(\phi_{0})}{(2l_{s}^{2})(4l^2 \beta)}
\frac{w^2}{3\frac{dY}{dt}}\nonumber\\
&=&
\frac{1}{\sqrt{3}\beta T}
\frac{w_{Y}^{2}}{3(\frac{dY}{dt})_{Y}}\nonumber\\
&=&\frac{1}{\sqrt{3}\beta T}
\frac{w_{Y}^{2}}{3(\frac{dY}{dt})_{Y}}
\eea
where,$\bar{\varrho}_{ind}$, $T^{'}_{ind}$ are factors independent of
$Y$ in $\bar{\varrho}$, $T^{'}$ respectively. On the otherhand,
we write $Y$-dependent factors of $w$ and $\frac{dY}{dt}$ 
as $w_{Y}$
and $(\frac{dY}{dt})_{Y}$. 
$\beta$ is the temperature at $Y=0$, or equivalently at $t=-\infty$. 
\par
Let us estimate
the holographic ratio in the zeroth order. This comes out as
\be
l_p^{2}(t)\frac{S}{A}=
\frac{1}{\sqrt{3}\beta T}
\ee
Now $\beta$ and $T$ are constants. Hence the holographic ratio also, in
the zeroth order, is constant, for all time. 
So entropy per unit Planck area of the horizon 
is also constant. It is given by the above
expression. This result is valid when the horizon is finite. Now 
the same solution describes the Universe in
the early phase i.e. when $t \rightarrow -\infty$
or, $a \rightarrow a_{0}$; the horizon tends to infinite then. Hence we 
infer, as a limiting procedure,
 that the entropy per unit Planck area of the
horizon is given by the expression(3.20) when 
the Universe is
flat. Since the area is tending to infinity,
entropy within the Hubble horizon 
also must be very large. As the Universe is cold at the
begining, it may appear contradictory. 
But the energy density per unit comoving voulme is
also constant. Hence entropy per comoving volume is constant. As the 
number of comoving volumes within the horizon 
tends to infinity, entropy 
within the Hubble horizon also tends to infinity.

This ratio has good physical implication. If this zeroth order ratio is one
(as is taken in time $-T$ in the paper\cite{venezia}), or,
at least bounded from above, then $\beta T = finite\quad number$.
Again in the PBB cosmology (i) $T$ which is the
 duration of dilaton driven phase
 is very large,
(ii) $\beta$, the temperature at the begining of the Universe, is very low. 
Hence
product, $\beta T$ is also a finite number. So two 
independent considerations lead to the same conclusion.
Recently Veneziano 
has shown$\cite{venezia}$  
that the ratio assumed of the order one,
explains the entropy budget of the Universe from the PBB cosmology
quite accurately upto some
numerical factors. 
Therefore it is quite reasonable to assume 
that the ratio is of the order 
one. In other words,
the Universe in the flat beginning seems to show holography as in
AdS spaces \cite{leny}.
\section{Holographic ratio near the remote past}
After discussing the features associated with the zeroth 
order holographic ratio
 being bounded let us see how the ratio changes
with time as the Universe evolves from the remote past. The evolution
will depend on the number distribution of strings at the remote past. 
As a result the holographic ratio will also be getting modified 
differentially.
Now the
holographic ratio
$l_{p}^{2}\frac{S}{A}$ is roughly 
$e^{\phi} \frac{D}{\sqrt{g}}$ . As the
Universe evolves the first factor increases whereas the 
second factor decreases. Hence the holographic ratio will increase or
decrease depending on whether the relative increament of the first
factor is more or less compared to relative decreament of the second 
factor. We will see in our study below that in the case of power law 
distribution the scale factor dominates whereas in the case of exponential
distribution it is the coupling constant which initially 
dominates, though very weakly, for a 
short interval of time before being overtaken by the scale factor.
\subsection{Power Law Distribution}
Let us now first consider the power law distribution,
$n_{-\infty}(L)=\Lambda^{2}L^{-3}$\cite{NET}. Then equation(3.5) becomes
\be
\pa_{Y}{\tilde{E}_{u}}-\tilde{E}_{u}=\Lambda^{2}\pa_{Y}{D^{-1}}
\ee
and on integration we get,
\be
\frac{\tilde{E}_{u}}{\Lambda^{2}}=H+e^{Y}\int_{0}^{Y}He^{-Y}
\ee
On the otherhand, if we directly evaluate eqn.(3.2) we arrive at,
\be
\frac{\tilde{E}_{s}}{\Lambda^{2}}=\frac{1}{l_{s}^{2}}-H
\ee
Hence, in this distribution, the expression 
for $\gamma$ takes the following form,
\be
\gamma=-\frac{1}{3}\frac{H+e^{Y}\int_{0}^{Y}He^{-Y}}
        {\frac{1}{l_{s}}+e^{Y}\int_{0}^{Y}He^{-Y}}
\ee
where,
$H$ is the Hubble parameter of the Universe given by
\be
H = \frac{dY}{dt}
\ee

Now from eqn.s(3.12)-(3.15) we get
\bea
\frac{dY}{dt}=\frac{1}{\sqrt{3}T}(cosh{\sqrt{3}Y}-1)\\
\frac{dx}{dY}=\frac{3}{2}\Gamma^{0}(cosech{\frac{\sqrt{3}}{2}Y})^{2}
\eea
where we have used the zeroth order expression for $w$,
\bea
w=\frac{4l}{T}e^{-\phi_{0}}sinh{\frac{\sqrt{3}}{2}Y}\nonumber
\eea
and then substituting eqn.(4.6) in eqn.(4.4), we get $\gamma$ to 
order $\epsilon$ as,
\be
\gamma=-\frac{\epsilon}{2}[cosh{\sqrt{3}}Y-e^{Y} 
+\frac{1}{\sqrt{3}}sinh{\sqrt{3}Y}]+o(\epsilon)^{2}
\ee
Here, integration of the R.H.S. of the expression (3.17) for $\Gamma$ is not
possible if we keep terms upto all orders in $Y$. So to start with, 
in the first step of iteration, we keep terms upto $Y^{2}$, for the
sake of simplicity and we get,
\be
\Gamma=\Gamma^{0} [1-\frac{\epsilon}{2}Y+o(Y)^{3}o(\epsilon)^{2}]
\ee
where, $\epsilon=\frac{l_{s}}{\sqrt{3}T}$.
Now, in the remote past $Y$ is small. 
So whether $T$ is large or small, correction
term in the above expression is small compared to one. 
Zeroth order solution acts as good perturbative vaccuum near
the remote past. Moreover, if $T$ is large, zeroth order solution acts as 
good perturbative vaccuum for all $Y$\footnote{By all $Y$ we mean that 
for all $Y$ for which power series converges}.

To carry on iteration, let us assume the $\Gamma$ obtained in eqn.(4.9)
to hold good for all orders in $Y$.
We put this $\Gamma$ in the background equation to get 
$w$ with $o(\epsilon)$ correction, 
\be
w= \frac{4l}{T}e^{-\phi_{0}}
(1+\frac{\epsilon}{2}Y)sinh{\frac{\sqrt{3}}{2}Y}
\ee
Consequently, 
\bea
\frac{dY}{dt}&=&\frac{1}{\sqrt{3}T}(cosh{\sqrt{3}}Y-1)\\
\frac{dx}{dY}&=&\frac{3}{2}\Gamma^{0}(cosech{\frac{\sqrt{3}}{2}Y})^{2}
                     [1-\epsilon Y+\frac{\epsilon}{\sqrt{3}}sinh{\sqrt{3}Y}]
\eea
Then, the model which takes care of string equations of
motion, gives
\be
\gamma=-\frac{\epsilon}{2}\frac{cosh{\sqrt{3}Y}-e^{Y} 
+\frac{1}{\sqrt{3}}sinh{\sqrt{3}Y}}
{1+\frac{\epsilon}{2}[\sqrt{3}sinh{\sqrt{3}Y}+cosh{\sqrt{3}Y}+2-3e^{Y}]}
\ee
Let us determine $w$ to order $\epsilon^{2}$.
Note that when we keep terms upto $Y^{4}$
the expression for $\gamma$ does not come with $\epsilon^{2}$.
Therefore, retaining terms upto $Y^5$ 
we find,
\be
\Gamma=\Gamma^{0}[1-\epsilon (Y+\frac{Y^2}{6} +\frac{Y^3}{36}
-\frac{Y^4}{240}-\frac{7 Y^5}{3600})+\frac{\epsilon^{2}}{40}Y^{5}
+o(Y)^{6}o(\epsilon)^{3}]
\ee
Let us use the expression for $\Gamma$ in the background 
equations of motion (3.12)-(3.14) to find the
$w$ and the corresponding fields upto $\epsilon^{2}$. 
Assuming that this 
 $\Gamma$ holds for all order in $Y$ we get the
expression for $w$ as in below:
\bea
w&=& \frac{4l}{T}e^{-\phi_{0}}\nonumber\\
&&[[1+\frac{\epsilon}{2}Y]sinh{\frac{\sqrt{3}}{2}Y}
+\epsilon[\frac{\sqrt{3}}{2}cosh{\frac{\sqrt{3}}{2}Y}
(-\frac{Y}{10}-\frac{2Y^2}{135}+\frac{Y^3}{180}+
\frac{7Y^4}{2160})\nonumber\\
&&+sinh{\frac{\sqrt{3}}{2}Y}(\frac{1}{10}+\frac{2}{135}Y+
\frac{3}{40}Y^2+\frac{1}{135}Y^3-\frac{1}{480}Y^4-
\frac{7}{7200}Y^5)]\nonumber\\
&&+\epsilon^{2}[-\frac{\sqrt{3}}{2}cosh{\frac{\sqrt{3}}{2}Y}
(\frac{119303}{437400}Y
+\frac{1}{324}Y^2+\frac{52883}{874800}Y^3-
\frac{403}{9720}Y^4\nonumber\\
&&-\frac{6997}{5832000}Y^5-
\frac{953}{3888000}Y^6 -\frac{677}{27216000}Y^7+
\frac{7}{1296000}Y^8\nonumber\\ 
&&+\frac{49}{34992000}Y^9 )+
 sinh{\frac{\sqrt{3}}{2}Y} (\frac{119303}{437400}
+ \frac{1}{324}Y + \frac{269423}{583200}Y^2\nonumber\\
&&+\frac{829}{19440}Y^3+\frac{68243}{2332800}Y^4
-\frac{16301}{1296000}Y^5-
\frac{14111}{23328}Y^{6}-\frac{11}{108000}Y^{7}\nonumber\\
&&-\frac{31}{3456000}Y^{8}+
\frac{7}{3456000}Y^{9}+
\frac{49}{103680000}Y^{10} )]
\eea
We can use this result to get $\gamma$ and $\Gamma$ for
third order or, upto $\epsilon^3$ and higher order in $Y$.
In other words, this procedure, in principle, could
be used to determine 
higher order terms in $\epsilon$.
 However, we compute terms upto
$\epsilon^{2}$.
Now,
we can use the 
first and second order expressions for $w$
to estimate the holographic ratio.
\par
Let us evaluate the holographic ratio
for the first order i.e. including the $\epsilon$ correction
\be
l_p^{2}(t)\frac{S}{A}=
\frac{1}{\sqrt{3}\beta T}(1+\epsilon Y)
\ee
The holographic ratio in the second 
order(i.e. when we include terms upto $\epsilon^{2}$) is
\bea
l_p^{2}(t)\frac{S}{A}&=&
\frac{1}{\sqrt{3}\beta T}\nonumber\\
&&[1-\epsilon(Y+\frac{Y^2}{6}+\frac{Y^3}{36}
-\frac{Y^4}{240}-
\frac{7Y^5}{3600}\nonumber\\
&&-\frac{-720-240Y-60Y^2 +12Y^3 
+7Y^4}{360\sqrt{3}}sinh{\sqrt{3}Y})\nonumber\\
&&-\epsilon^{2}[-\frac{139}{405}-\frac{521Y}{2430}
-\frac{5233Y^2}{4860}
-\frac{16297Y^3}{48600}\nonumber\\
&&-\frac{5429Y^4}{72900}
-\frac{1501Y^5}{64800}+\frac{1877Y^6}{388800}\nonumber\\
&&+\frac{Y^7}{1296}+\frac{23 Y^8}{388800}-\frac{7Y^9}{432000}
-\frac{49Y^{10}}{12960000}\nonumber\\
&&-\frac{139(-720-240Y-60Y^2 +12Y^3 +7Y^4)
cosh{\sqrt{3}Y}}{145800}\nonumber\\
&&-(1601280+1000320 Y+358080Y^2 + 9312Y^3\nonumber\\
&&-41344Y^4 - 12888Y^5 - 1500Y^6 + 504Y^7 + 147Y^8)
\frac{cosh{2\sqrt{3}Y}}{4665600}\nonumber\\
&&+(-2Y-Y^2 -\frac{Y^3}{3}-\frac{55Y^4}{216}+
\frac{73Y^5}{1800}+
\frac{133Y^6}{14400}\nonumber\\
&&+\frac{47Y^7}{43200}-\frac{7Y^8}{32000}-
\frac{49Y^9}{864000})  
\frac{sinh{\sqrt{3}Y}}{\sqrt{3}}\nonumber\\
&&+(\frac{1}{10}+\frac{17Y}{270}+
\frac{Y^2}{648}
-\frac{287Y^3}{16200}\nonumber\\
&&-\frac{31Y^4}{4320}-\frac{53Y^5}{48600}+
\frac{49Y^6}{129600}
+\frac{49Y^7}{388800})\frac{sinh{2\sqrt{3}Y}}{\sqrt{3}}]] 
\eea
We observe  
that coefficient of $\epsilon$ in the above equation(4.17)
is of opposite sign compared to that of first order eqn.(4.16). 
This sign difference arises due to the following reason.
In obtaining the expression(4.9) for $\Gamma$ we have kept terms,
upto $Y^2$. But
while calculating $w$, eqn.(4.10), we have kept terms upto
all orders in $Y$. 
Then we have used this expression (4.10) 
for $w$ in the holographic formula. 
It gives rise to the
apparent discrepency. 
If we had strictly kept terms upto $Y^2$, coefficient of 
$\epsilon$ would have been absent in the first order
holographic ratio.  
Again in the second order
formula(4.17) there is no term in 
the coefficient of $\epsilon$ upto $Y^{2}$.
Hence the second order result 
is consistent with the first order one.
Moreover, if we had kept terms in the expression for $w$ upto
$Y^5$, strictly, the 
holographic ratio with $\epsilon^{2}$ correction
would have contained terms upto $Y^3$.
Now in the expression for the holographic ratio(4.17), 
coefficient of $\epsilon$ decreases and that of $\epsilon^{2}$
increases with $Y$, if terms upto $Y^3$
are considered. This feature will be manifest from the plots below. 
Again as a next step, if we go on to consider 
the third order in $\epsilon$,
 to start with we have to keep terms higher order than
$Y^{5}$ in the expression for $\Gamma$. And this will not
change the coefficients of terms upto $Y^{5}$ in the 
expression for $w$ and the coefficients of terms upto $Y^{3}$
in the expression for holographic ratio. 
Hence, we conclude, that as the Universe
evolves form $t=-\infty$, the holographic ratio decreases,  
at least upto the time terms from $Y^{6}$ onwards
are negligible.\\
The plots of the coefficients
of $\epsilon$ and $\epsilon^{2}$ in the holographic ratio, 
for small values of $Y$, are shown in Fig. 2 and 3 respectively.
\begin{figure}[h]
\centerline{\leavevmode\epsfysize=5truecm \epsfbox{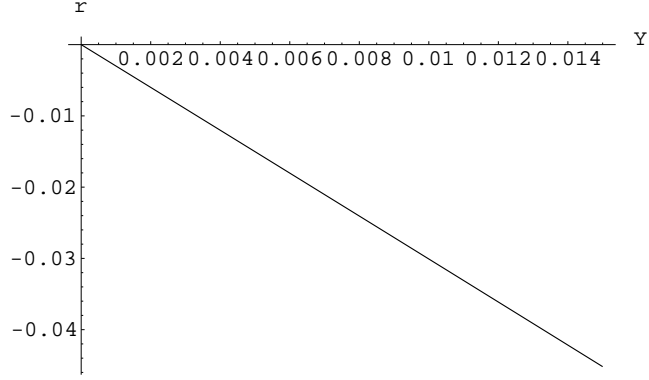}}
\vspace{0.2in}
\caption{ The plot of $Y-dependent$ coefficient 
of $\epsilon$ in $r=l_{P}^{2}\frac{S}{A}$
against $Y$.}
\label{a}
\end{figure}
\begin{figure}[h]
\centerline{\leavevmode\epsfysize=5truecm \epsfbox{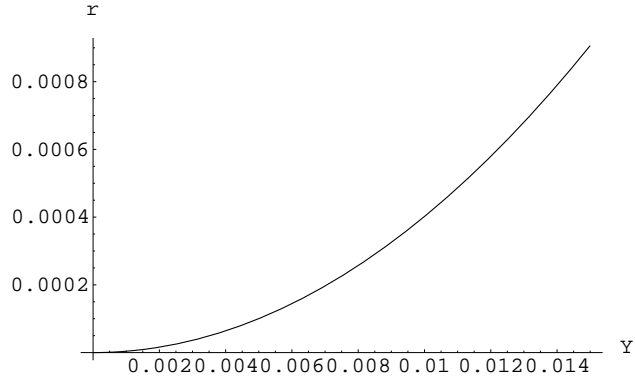}}
\vspace{0.2in}
\caption{ The plot of $Y-dependent$ coefficient of $\epsilon^{2}$ in 
$r=l_{P}^{2}\frac{S}{A}$ against $Y$.}
\label{a}
\end{figure}
We observe from the plots that the coefficient of $\epsilon$ decreases 
whereas the coefficient of $\epsilon^{2}$ increases. Note that in both
the cases the coefficients are indeed small.
As a result the
net holographic ratio decreases as stated in the previous paragraph.

\subsection{Exponential distribution}
After discussing all about the power law distribution, let us go over
to the second case, namely exponential one.
The differential equation satisfied by $\tilde{E}_{u}$ for the exponential
distribution takes the form,
\be
\pa_{Y}{\tilde{E}_{u}}-\tilde{E}_{u}=
-N'\pa_{Y}{D} \frac{D}{L_{0}}exp(-\frac{D}{L_0})   
\ee
which on integration gives,
\be
\tilde{E}_{u}=N'L_{0}[\frac{D}{L_{0}}exp(-\frac{D}{L_0})-exp(Y)\int_{0}^{Y}
\pa_{Y}{(exp(-Y)\frac{D}{L_{0}})}exp(-\frac{D}{L_0})dY]
\ee
Since, $H=D^{-1}=\frac{dY}{dt}=\frac{\sqrt{3}}{2T}Y^2$ upto order of $Y^2$ 
we get, 
\be
\gamma=-\frac{1}{3}\frac{1}{fY^{2}}exp(-\frac{1}{fY^{2}})
\ee
Use of the eqn.(3.17) leads to
\be
\Gamma=\Gamma^{0}
[1-\frac{1}{3}(\frac{exp(-\frac{1}{fY^{2}})}{Y}-
\frac{\sqrt{\pi f}}{2}(Erf(\frac{1}{\sqrt{f}Y})-1))]
\ee
where, $Erf(\frac{1}{\sqrt{f}Y})$ is the error function\cite{grad}.
Now to iterate we assume that the expression(4.21) for
 $\Gamma$ to be correct to all order
in $Y$. We use this $\Gamma$ in the differential equation(3.14) and
obtain $w$. Here, we keep terms upto $Y^{5}$ in the expression for $w$ 
and get
\be
w=\frac{4l}{T}exp(-\phi_0)[sinh(\frac{\sqrt{3}}{2}Y)+c(Y)]
\ee
where,
$c(Y)$ is given by
\bea
c(Y)&=&\frac{1}{3}sinh(\frac{\sqrt{3}}{2}Y)\left[(\frac{1}{Y}+\frac{3Y}{8f})
exp(-\frac{1}{fY^{2}})
-\sqrt{f\pi}(Erf(\frac{1}{\sqrt{f}Y})-1)
(1+\frac{3}{4f}-\frac{3}{8f^2})\right]\nonumber\\
&&-\frac{1}{3}cosh(\frac{\sqrt{3}}{2}Y)
\left[\frac{\sqrt{3}}{2}
(1+\frac{3}{40}\frac{Y^2}{f})exp(-\frac{1}{fY^{2}})
-\frac{\sqrt{3}}{4f}
(1-\frac{3}{40f})ExpIntegralEi(-\frac{1}{fY^{2}})\right]\nonumber\\
&&
\eea
where, $ExpIntegralEi(-\frac{1}{fY^{2}})$ is the exponential integral
function \cite{grad}.

Then the holographic ratio becomes,
\be
l_p^2(t)\frac{S}{A}=
\frac{1}{\sqrt{3}\beta T}[1+COR(Y)]
\ee
with $COR(Y)$ is given by,
\bea
COR(Y)&=&exp(-\frac{1}{fY^{2}})[\frac{8}{3\sqrt{3} f^2 Y^{7}}+
\frac{2}{\sqrt{3}f^2 Y^{5}}-\frac{4}{\sqrt{3}fY^{5}}-\nonumber\\
&&\frac{4}{3\sqrt{3}Y^{3}}+\frac{1}{2\sqrt{3} f^2 Y^{3}}+
\frac{1}{3fY^{3}}-\frac{5}{\sqrt{3}fY^{3}}+\frac{1}{2Y}-\nonumber\\
&&\frac{1}{2\sqrt{3}Y}+\frac{1}{80 f^2 Y}-\frac{3\sqrt{3}}{80 f^2 Y}+
\frac{1}{6fY}-\frac{\sqrt{3}}{2fY}+\frac{Y}{40f}\nonumber\\
&&-\frac{3\sqrt{3}Y}{40f}+
\frac{1}{2}coth(\frac{\sqrt{3}}{2}Y)(1-\frac{\sqrt{3}}{4}+
\frac{3Y^2}{40f}-\frac{3\sqrt{3}Y^2}{160f})\nonumber\\
&&-cosh(\frac{3\sqrt{3}}{2}Y)cosech(\frac{\sqrt{3}}{2}Y)
(\frac{1}{8\sqrt{3}}+\frac{\sqrt{3}Y^2}{320f})]\nonumber\\
&&-exp(-\frac{1}{f Y^{2}})[-\frac{1}{4\sqrt{3}}
+\frac{3}{40f}+\frac{1}{8\sqrt{3}f}-\frac{4}{3 f^2 Y^{6}}\nonumber\\
&&-\frac{1}{20 f^3 Y^{4}}-
\frac{2}{3 f^2 Y^{4}}
+\frac{2}{fY^{4}}+\frac{2}{3\sqrt{3}fY^{4}}+
\frac{1}{3Y^2}+\frac{1}{3\sqrt{3}Y^2}\nonumber\\
&&-\frac{3}{40 f^2 Y^2}
+\frac{5}{6fY^2}+\frac{1}{2\sqrt{3}fY^2}
-\frac{\sqrt{3}Y^2}{160f}]sinh(\sqrt{3}Y)\nonumber\\
&&-cosh(\sqrt{3}Y)[exp(-\frac{1}{fY^{2}})(\frac{8}{3\sqrt{3} f^2 Y^7}
+\frac{2}{\sqrt{3} f^2 Y^5}-\frac{4}{\sqrt{3}f Y^5}\nonumber\\
&&-\frac{4}{3\sqrt{3} Y^3}+\frac{1}{2\sqrt{3} f^2 Y^3}
-\frac{1}{3fY^3}-\frac{\sqrt{3}}{fY^3}+\frac{1}{6Y}\nonumber\\
&&-\frac{1}{2\sqrt{3}Y}-\frac{1}{80 f^2 Y}
-\frac{\sqrt{3}}{80 f^2 Y}-\frac{1}{6fY}-\frac{1}{2\sqrt{3}fY}\nonumber\\
&&-\frac{Y}{40f}-\frac{\sqrt{3}Y}{40f})
+\frac{\sqrt{f\pi}}{12}(1-\sqrt{3})(1-Erf(\frac{1}{\sqrt{f}Y}))]\nonumber\\
&&+(\frac{(3-\sqrt{3})(3-40f)}
{480f^2}coth(\frac{\sqrt{3}}{2}Y) 
ExpIntegralEi(-\frac{1}{f Y^{2}}))
\eea
To see how the holographic ratio
 changes with $Y$, we plot the
correction to ratio
against $Y$ for f equal to 10 in figure 4.:
\begin{figure}[h]
\centerline{\leavevmode\epsfysize=5truecm \epsfbox{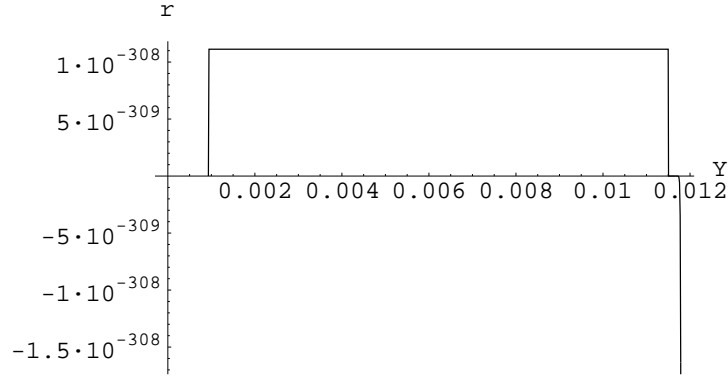}}
\vspace{0.2in}
\caption{ The plot of correction to holographic 
ratio $l_{P}^{2}\frac{S}{A}$ denoted
as $r$, against $Y$ for exponential distribution with $L_{0}=fT$ and $f=10$.}
\label{a}
\end{figure}
We note that the holographic ratio, in this case, 
deviates slightly upward from the zeroth order result.
Let us check that this deviation is not spurious. For this purpose
we plot the correction to holographic
ratio for $w$ upto $Y^{2}$ order also.
We get an identical looking curve.
We recall that we have kept terms upto $Y^{2}$
in the expression for $H$. This means 
that the region of $Y$ taken in the plot is the region where higher order
terms are negligible compared to $Y^{2}$. Hence this little upward deviation
from the zeroth order result will remain if we 
go to higher order in $Y$.
Actually this little upward deviation is generic of exponential
distribution. Again, to come to this
conclusion, we plot the ratio for $f=1$. This is shown in the figure 5:
\begin{figure}[h]
\centerline{\leavevmode\epsfysize=5truecm \epsfbox{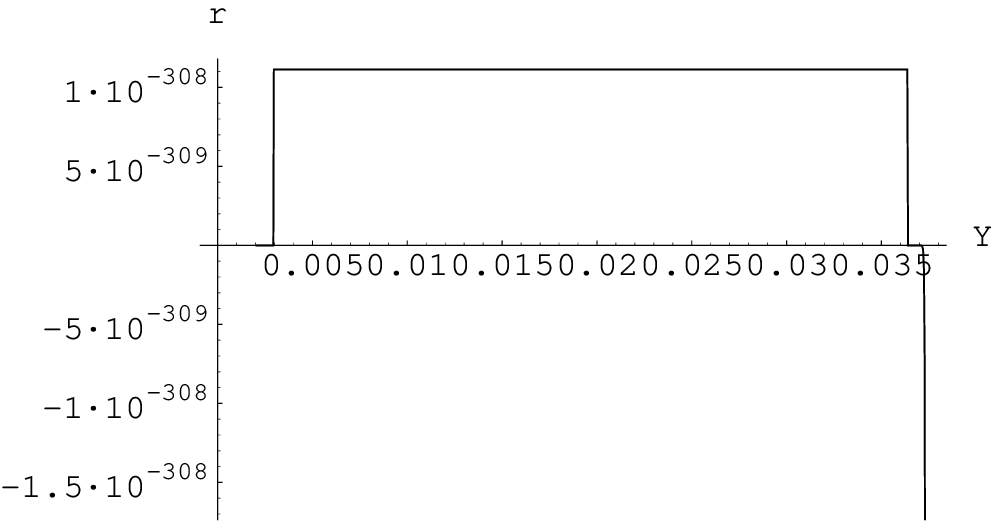}}
\vspace{0.2in}
\caption{ The plot of correction to holographic 
ratio $l_{P}^{2}\frac{S}{A}$ denoted
as $r$ against $Y$ for exponential distribution with $L_{0}=fT$ anf $f=1$.}
\label{a}
\end{figure}
We see that the domain of the 
initial plataue or, rectangular shaped region
 in the above figure, $FIG. 5$, shifts with respect to the previous
plot, $FIG. 4$, to slightly larger value 
of $Y$. Hence we infer that if we go on decreasing $f$ to $\epsilon$
this nature of the curve will remain, only the
domain of the curve will shift to larger value of $Y$. 
This plataue occurs due to first order correction. 
So the initial rise will remain if we include 
the higher order corrections also.
Note the magnitude of smallness in the initial rise in the holographic
ratio. We mention in passing that for $\gamma=-\frac{1}{3}$ solution to be
valid, $f=\frac{L_{0}}{T}=\frac{t_{cl}}{T}$ has to be much greater
than $\frac{l_{s}}{T}$ but less than one, as $t_{cl}<T$.

Uptill now we were concentrating on how the holographic 
ratio evolves over small time interval near the remote 
past. In this regime, only few strings cross the horizon. 
Perturbative techniques are useful also. But we don't
have an easy way to study the holographic behaviour when,
arbitrary number of strings say, half
of the total number of strings are more than the size of the horizon. Obvious 
possibility is to go to the other extreme
i.e. when all the strings became nondynamical.
This is what we are going to do in the next section.
\section{Holographic ratio for $\gamma=-\frac{1}{3}$}
The holographic ratio is given by (for
$\gamma=-\frac{1}{3}$ or, $t\sim -t_{cl}$)
\be
l_p^{2}(t)\frac{S}{A}=
\frac{1}{3\beta_{t_{cl}}t_{0}}
(\frac{-t}{t_{0}})^{-\frac{1}{2}}
\ee
where, $\beta_{t_{cl}}$ is the temperature at the time
$t=-t_{0}$.
So the holographic ratio at the time $t=-t_{cl}$ is given by  
$\frac{1}{3\beta_{t_{cl}} (-t_{cl})^{\frac{1}{2}}(t_{0})^{\frac{1}{2}}}$.
Hence the ratio of the holographic ratios, denoted by $R_{H}$, at the time 
$-t_{cl}$ and $-\infty$ is
\be
\frac{R_{H}(-t_{cl})}{R_{H}(-\infty)}=
\frac{3\sqrt{3}}{4}\frac{T}{(-t_{cl})^{\frac{1}{2}}}
(\frac{e^{\phi_{0}}}{3l})^{\frac{1}{2}}
\ee
where we have used the constancy of comoving volume entropy.
Now assuming that the holographic ratio is one at the remote past we
find that the ratio at the time $-t_{cl}$ is also one provided the 
parameters of the theory $\phi_{0}$ and $l$ are
constrained by the relation
$(\frac{e^{\phi_{0}}}{3l})^{-\frac{1}{2}}=
\frac{3\sqrt{3}}{4}\frac{T}{(-t_{cl})^{\frac{1}{2}}}$.
Note that $T$ is larger than $|t_{cl}|$ (see the Fig.1).

\centerline{\bf{Comparison with the Einstein frame ratio}}

Let us study another feature of the holographic ratio. This is whether 
the holographic ratio is the 
same in both the string and Einstein
frames. In the PBB cosmology, qualitative features of the Universe
like blue shift, shrinking of the horizon
etc. \cite{MPL}
are same in both the frames. So the question whether similarly
 the holographic
ratio is equal in the two frames motivates us to study the
corresponding ratio in the Einstein frame.
The metrics in the Einstein frame and the string frame 
are related by the following conformal
transformation
\be
8\pi g^{s}_{\mu\nu}=e^{\phi-\Phi_{0}}g^{E}_{\mu\nu}
\ee
$\Phi_{0}$ is the present day value of the dilaton and we set
$16 \pi \l_{s}^{2}e^{\Phi_{0}}=l_p^{2}$.
Then if we remain in the synchronous gauge i.e. if we set
\be
ds_{E}^{2}=dt_{E}^{2}-a_{E}^{2}(dx^{i})^{2}
\ee
we get
\bea
dt_{E}&=&\sqrt{8\pi}e^{-\frac{1}{2}(\phi-\Phi_{0})}dt\\
a_{E}(t_{E})&=&\sqrt{8\pi}e^{-\frac{1}{2}(\phi-\Phi_{0})}a(t)\\
\varrho^{s}_{E}(t_{E})&=&\frac{e^{2(\phi-\Phi_{0})}}{(8\pi)^{2}}
\varrho^{s}(t)
\eea
as,
$(T_{\mu}^{\mu})_{E}=\frac{\sqrt{g_{s}}}{\sqrt{g_{E}}}(T_{\mu}^{\mu})$
\cite{MPL}.
In the Einstein frame the equations of motion satisfied by the fields
are
\bea
R_{\mu}^{\nu} -\frac{1}{2}R g_{\mu}^{\nu}&=&\frac{l_{p}^{2}}{2}T_{\mu}^{\nu}\cr
\nabla_{\mu}\nabla^{\mu}\phi&=&\frac{l_{p}^{2}}{2}(T^{S})_{\mu}^{\mu}
\eea
where, in the above,
$T_{\mu}^{\nu}=(T^{D})_{\mu}^{\nu}+(T^{S})_{\mu}^{\nu}$ 
and $(T^{D})_{\mu}^{\nu}$ is the stress-energy tensor corresponding 
to the dilaton. 

These equations of motions
have been derived from an action $S^{E}$ which in turn
has been obtained from the effective 
action in the string frame by conformal transformation and is given as
\be
\hbar^{-1}S^{E}=-\frac{1}{l_{p}^{2}}\int{ d^{4}x \sqrt{-g_{E}}
       (R-\frac{1}{2}\partial_{\mu}\phi\partial^{\mu}\phi)}\\
      +S_{\sigma}
\ee
where,
\be
S_{\sigma}=\frac{1}{4\pi\alpha^{'}}
           \int{d^{2}\sigma \partial_{\alpha}X^{\mu}
           \partial^{\alpha}X^{\nu}g_{\mu\nu}}e^{\phi}.
\ee
After this short preliminary about the conversion of string frame 
to Einstein frame in $(1+3)$-dimension, let us 
proceed to compute the ratio in the Einstein frame corresponding to
$\gamma=-\frac{1}{3}$ in the string frame.
We note that in the Einstein frame\cite{DPISC},
\bea
a_{E}(t_{E})&=&\frac{\sqrt{8\pi}}
{e^{\frac{\phi_{0}^{'}}{2}}}
              (\frac{-t_{E}}{t_{0E}})^{\frac{2}{5}}\\
\phi(t_{E})&=&\phi_{0n}-\frac{6}{5}\ln(\frac{-t_{E}}{t_{0E}})\\
\varrho^{s}(t_{E})&=&\frac{6}{25 L_{p}^{2}}\frac{1}{(-t_{E})^{2}}
\eea
where, in the above, $\varrho^{s}(t_{E})$ is the energy density of the
string source in the Einstein frame and $L_{p}$ is the present
day value of the Planck length. Again in the Einstein frame,
the dilaton also contributes to the energy density and the pressure.
 Consequently, the net
eneregy density and the pressure in the Einstein frame are
\be
\varrho=\frac{24}{25 L_{p}^{2}}\frac{1}{(-t_{E})^{2}},\quad
p=\frac{16}{25 L_{p}^{2}}\frac{1}{(-t_{E})^{2}}
\ee
The effective equation of state is
\be
p=\frac{2}{3}\varrho
\ee
The holographic ratio turns out to be
\bea
L_{p}^{2}\frac{S}{A}&=&L_{p}^{2}\frac{S^c}{\sqrt{g}}\frac{D}{3}\nonumber\\
&=&L_{p}^{2}\frac{1}{\sqrt{g}}\frac{\bar{\varrho}(1+\gamma)}{T^{'}}
\frac{1}{3|\frac{\dot{a}}{a}|}\nonumber\\
&=&\frac{4}{3 \beta_{t_{cl}}^{E}}
\frac{(-t_{E})^{-\frac{1}{5}}}{(t_{0E})^{\frac{4}{5}}}
\eea
where, $\beta_{t_{cl}}^{E}$ is the temperature at the time $t_{0E}$.
Now, the string frame time and the Einstein frame time are related by
\bea
t_{0}&=&\frac{5}{2\sqrt{8\pi}}e^{\frac{\phi_{0}^{'}}{2}}t_{0E}\nonumber\\
-t&=&\frac{5}{2\sqrt{8\pi}}e^{\frac{\phi_{0}^{'}}{2}}(t_{0E})^{\frac{3}{5}}
       (-t_{E})^{\frac{2}{5}}
\eea
we get,
\be
\frac{(-t_{E})^{-\frac{1}{5}}}
{(t_{0E})^{\frac{4}{5}}}=\frac{5}
{2\sqrt{8\pi}}e^{\frac{\phi_{0}^{'}}{2}}
\frac{1}{t_{0}}(\frac{-t_{}}{t_{0}})^{-\frac{1}{2}}
\ee
As a result the ratio, in the two frames eqn.(5.1) and eqn.(5.16)
 have exactly the same time dependences.
Again the entropy per comoving volume is the same in both the 
string and the Einstein frames. Hence
\be
\beta_{t_{cl}}^{E}=\frac{10}
{(8\pi)^{\frac{1}{2}}}e^{\frac{\phi_{0}^{'}}{2}}
\beta_{t_{cl}}
\ee
where, $\phi_{0}^{'}=\phi_{0n}-\Phi_{0}$ and $\Phi_{0}$ is the present
day value of the dilaton.
Therefore the holographic ratios in both the frames are exactly the same. 
\par
Uptill now we have been in $(1+3)$-dimension, came across nice features 
like constancy of holographic ratio in the remote past, 
identical ratio in the string and Einstein frames 
in the recent past,i.e. $|t|=t_{cl}=t_{c}$. Let us now move
on to diverse dimensions 
looking for the similar interesting features.
\section{Holographic ratio in diverse dimension}
Here we write down the string effective action in $(1+d)$-dimension and 
the equations of motions derived from it. 
The effective action in $(1+d)$-dimension is
\be
\hbar^{-1}S^{s}=-\frac{1}{2 l_{s}^{d-1}}
      \int{d^{d+1}x \sqrt{-g} e^{-\phi}(R+
      \partial_{\mu}\phi \partial^{\mu}\phi)}
      +S_{\sigma}
\ee
The resulting equations of motions \cite{PBBISC} are
\bea
\dot{\bar{\phi}}^{2}-2\ddot{\bar{\phi}}+ mH^{2}+ nF^{2}&=&0\nonumber\\
\dot{\bar{\phi}}^{2}-mH^{2}-nF^{2}&=&2l_{s}^{d-1}\bar{\varrho}
e^{\bar{\phi}}\nonumber\\
2(\dot{H}-H\dot{\bar{\phi}})&=&2
\l_{s}^{d-1}\bar{p}e^{\bar{\phi}}\nonumber\\
2(\dot{F}-F\dot{\bar{\phi}})&=&2
\l_{s}^{d-1}\bar{q}e^{\bar{\phi}}
\eea
First let us keep in mind that we are writing $d=m+n$, 
where, (i) $m$ is the 
number of expanding spatial dimensions and (ii) $n$ is the number of 
contracting spatial dimensions in the string frame. 
Scale factor of a contracting dimension will be
denoted as $a_{con}$ and it will be just reciprocal of 
$a_{ex}$. This is the only posssibility compatible with non-zero
energy density \cite{PBBISC}.
Here, now onwards we will be
confining ourselves to consider mainly time dependences 
of quantities.
Moreover we will look into the following aspects, 
(i) whether the holographic ratio is
constant or, not in different dimensions in the remote past, 
(ii) whether the holographic ratios in the remote past are the 
same in both the string and the
Einstein frames in $(1+3)$ as well as in the 
diverse dimensions and 
(iii) how the time dependences of the holographic ratios in the recent past, 
in the string frame, go together with the corresponding ones in the 
Einstein frames in various dimensions.
\subsection{Remote Past}
Let us now start with the case when we have nine isotropically
 expanding dimensions.
Then we have\cite{NET}
after solving the above set(6.2) of equations
as in the $(1+3)$-dimensional case with $p=q=0$,  
\bea
a&=&a_{0}(1-\frac{2T}{t})^{\frac{1}{3}}\nonumber\\
e^{\bar{\phi}}&=&\frac{16 l^{2} e^{-\phi_{0}}}{|t(t-2T)|}\nonumber\\
\bar{\varrho}&=&\frac{e^{\phi_{0}}}{8 l^{2}l_{s}^{8}}
\eea
And we know \cite{dten} in the ten dimension, $Planck length$, is
given by,
\be
(l_{p}^{10})^{2}= g_{s}^{\frac{1}{2}}l_{s}^{2}=e^{\frac{\phi}{4}}l_{s}^{2}
\ee
So, the holographic ratio is given by,
\bea
(l_{p}^{10})^{8}\frac{S}{A}&=&l_{s}^{8}\frac{e^{\phi}}{\sqrt{g}}
\frac{\bar{\varrho}}{\beta_{9e}}\frac{V_{H}}{A_{H}}\nonumber\\
&=&\frac{3}{2k}\frac{1}{\beta_{9e}T}
\eea
where, $k$ is a numerical factor coming through
 $\frac{V_{H}}{A_{H}}=\frac{1}{k H}$
and $\beta_{9e}$ is the temperature in the remote past.
\par
Next let us consider the case when three dimensions are expanding
and six other dimensions are contracting. Then, observing that
equations of motions in terms of SFD
invariant variables remain same as in the previous case, we get
\bea
a_{ex}&=&a_{0ex}|1-\frac{2T}{t}|^{\frac{1}{3}}\nonumber\\
a_{con}&=&a_{ex}^{-1}\nonumber\\
e^{\bar{\phi}}&=&\frac{16 l^{2} e^{-\phi_{0}}}{|t(t-2T)|}\nonumber\\
\bar{\varrho}&=&\frac{e^{\phi_{0}}}{8 l^{2}l_{s}^{8}}
\eea
And 
$(l_{p}^{10})^{2}$ is written as in the nine expanding case. So, the
holographic ratio
\be
(l_{p}^{10})^{8}\frac{S}{A}=l_{s}^{8}\frac{e^{\phi}}{\sqrt{g}}
\frac{\bar{\varrho}}{\beta_{3e}}\frac{V_{H}}{A_{H}}
\ee
Again, the ratio of the Hubble volume to Hubble area is given by 
\bea
\frac{V_{H}}{A_{H}}
&=&\frac{\sqrt{\pi}}{36(\frac{\sqrt{\pi}}{2}|H| + \frac{16}{15}|F|)}\nonumber\\
&=&\frac{\sqrt{\pi}}{24}\frac{1}{\frac{\sqrt{\pi}}{2} + \frac{16}{15}}
\frac{t}{T}(t-2T)
\eea
where, $|H|$ and $|F|$ are the Hubble parameters of the expanding and
the contracting dimensions.
Consequently,
\be
(l_{p}^{10})^{8}\frac{S}{A}=\frac{\sqrt{\pi}}{12}
\frac{1}{\frac{\sqrt{\pi}}{2} + \frac{16}{15}}\frac{1}{\beta_{3e}T}
\ee
Therefore, 
we see, in both the cases as $t$ tends to $-\infty$, the holographic ratio
 becomes constant as in the $(1+3)$-dimension. 
Let us examine whether this happens in any arbitrary dimensions.

We note that for $(1+d)$-dimensional world,
\bea
a_{ex}&=&a_{0ex}(1-\frac{2T}{t})^{\frac{1}{\sqrt{d}}}\nonumber\\
e^{\bar{\phi}}&=&\frac{16 l^{2} e^{-\phi_{0}}}{|t(t-2T)|}\nonumber\\
\bar{\varrho}&=&\frac{e^{\phi_{0}}}{8 l^{2}l_{s}^{d-1}}
\eea
And as far as time dependence is concerned, in the large $|t|$ limit
(i.e. $|t|>>T$) we have
\bea
l_{p}^{d-1}(t) \frac{S}{A}&\sim& 
\frac{V_{H}}{A_{H}}\frac{e^{\phi}}{\sqrt{g}}\nonumber\\
&\sim&\frac{e^{\phi}}{\sqrt{g}}\frac{1}{|H|}\nonumber\\
&\sim& |t|^{0}
\eea

Again, in the $(1+d)$-dimension, the Einstein frame scale factor and time 
are related to the string frame ones\cite{MPL} as 
\bea
a_{E}(t_{E})&=&e^{-\frac{\phi}{d-1}} a(t)\nonumber\\
|t_{E}|&=& |t|^{\frac{d+1}{d-1}}
\eea
Thus the scale factor in the Einstein frame is, in the large 
$|t|$ limit, in terms of the string frame time is expressed as
\be
a_{E}(t_{E})= |t|^{\frac{2}{d-1}}
\ee
Note that in this limit the Universe is isotropic even if 
in the string frame it is mixed isotropic. In 
$(1+3)$-dimension, in particular, the equation of state is
\be
p=\frac{1}{3}\varrho
\ee
 Then we arrive at the following holographic ratio in the Einstein
frame, in the $(1+d)$-dimension
\be
L_{p}^{d-1}\frac{S}{A} \sim \frac{1}{\sqrt{g_{E}}}
|\frac{a}{\frac{da}{dt_{E}}}| \sim |t_{E}|^{\frac{2-d}{2}} 
          \sim |t|^{-1}
\ee
where, $|t_{E}|$ corresponds to the time in the Einstein frame.
So the time dependences of the holographic ratios
in the two frames in this large $|t|$ limit do not match
in any dimension. But we have seen that in the recent past the
ratio is same in both the frames in the four dimension. Therefore, let us
check, in the recent past, whether 
in arbitrary dimensions time dependences of the
holographic ratio match in both the frame.
\subsection{Recent Past}
Now in the $(1+d)$-dimensional spacetime near the time $-t_{cl}$, 
$\gamma=-\frac{1}{d}$ 
and the relevent field configurations, namely
 the scale factor of the expanding 
dimension and the SFD invariant dilaton\cite{MPL} are 
\bea
a({t})&=&|t|^{-\frac{2}{1+d}} \nonumber\\
\bar{\phi}&=&dln|a(t)|
\eea
As a result the holographic ratio in the string frame
becomes 
\be
l_{p}^{d-1}\frac{S}{A} \sim |t|^{\frac{1-d}{1+d}}
\ee
\newline
But now the Einstein frame time $t_{E}$ is related to the string frame
time by 
\be
|t_{E}|\sim |t|^{1+\frac{4m}{d^2 -1}}
\ee
and the scale factors in the Einstein frame, 
corresponding to the expanding and contracting
dimensions in the string frame, are
\bea
a_{Em}(t_{E})\sim |t_{E}|^{2\frac{m+1-n}{d^2 +4m -1}} 
\sim |t|^{-2\frac{n-m-1}{d^2 -1}}  \nonumber\\
a_{En}(t_{E})\sim |t_{E}|^{2\frac{2m-1+d}{d^2 +4m -1}}
\sim |t|^{2\frac{d-1+2m}{d^2 -1}}
\eea
Consequently, the corresponding ratio in the Einstein frame in terms 
of the string time is given by
\be
L_{p}^{d-1}\frac{S}{A} \sim
|t|^{\frac{1-d}{1+d}}
\ee
Hence in the recent past the holographic
ratios in the two frames have the same time dependences in all dimensions. 
We note here that time 
dependences of holographic ratio as in eqn.(6) was arrived
at\cite{bak} 
from different consideration.
 
\section{Discussion}
In this paper, we 
have set out to compute the holographic ratio. First, we
obtained the holographic ratio in the zeroth order and
discussed the implication of the ratio to be bounded. Then 
for the power law distribution,
we obtained the 
correction to the ratio
in the first and second order in $\epsilon$ following an iteration
procedure. 
We found that the correction depletes the ratio, albeit by small amount.
For the exponential distribution, the
correction increases the value of the ratio by very small 
amount ($10^{-308}$),
towards the beginning of evolution, over a small range of time.
This is due to the domination of coupling constant over scale factor 
during that period.
Again, we computed the ratio in the other extreme, when all the strings 
crossed the horizon, assuming the mean length of the classical strings 
is much larger than $l_{s}$. We then showed that the holographic ratio,
has the same time dependences in both the string and Einstein frames 
in this limit, and taking care of all the factors in one particular 
case, namely
in the four dimension we found that the ratio is identical.
This gives another 
evidence that qualitatively
the two frames are equivalent in the PBB cosmology and Hubble horizon
is a good choice in both the frames.
On the other hand, in the remote past the ratio
in the string frame is constant 
in all dimensions.
However,
in the Einstein frame, it goes as inverse of the string time.
The ratio is different in the two frames.
We have checked that solutions obtained by conformal trasformation
do satisfy the equations of motions in the Einstein frame in this
limit when coupling constant goes to zero.
We may mention that if one follows the proposal of ref\cite{bak},
then also one gets the corresponding holographic ratio which is 
inversely proportional to the string time.
Thus we find that the holographic ratio in the remote past is not on
similar footing as other physical quantities. 
We have studied the holographic behaviour in the presence of 
string sources almost in two extreme cases in $(1+3)$-dimensions.
It is possible to generalise the exercise of getting 
 the correction by iterative procedure to
$(1+d)$-dimension.
Again, we applied
pertubative method to go away from the extremality ($\gamma=0$). 
It will be
very interesting to study the holographic ratio explicitly in
the intermediate region.
It is also desirable to see how the zeroth order holographic ratio 
gets modified in presence of higher dimensional branes.
Moreover it is very much of relevence to 
ask about the mechanism of holography in this case.

\centerline{Acknowledgement}

We have benifitted from discussions with 
Professors R. Brustein, M. Gasperini 
and especially with Professor A. Sen.
We are grateful to Professor G. Veneziano 
for many valuable discussions
at various stages of the work
and this investigation is outcome of his suggestions
to explore consequences of holography
in presence of the string sources.
We are thankful to D. Goswami for drawing the first figure on behalf of us.

\end{document}